# THE SHADOW OF LIGHT: FURTHER EXPERIMENTAL EVIDENCES

F. Cardone[a], R. Mignani[b,c,*], W. Perconti[d], A. Petrucci[b], R. Scrimaglio[d]

[a] Istituto per lo Studio dei Materiali Nanostrutturati (ISMN-CNR), via dei Taurini 19, 00185 Roma, Italy

[b] Dipartimento di Fisica "E. Amaldi", Università degli Studi "Roma Tre", via della Vasca Navale. 84, 00146 Roma, Italy

[c] INFN, Sezione di Roma III

[d] Dipartimento di Fisica, Università degli Studi de L'Aquila, via Vetoio 1, 67010 Coppito, L'Aquila , Italy

*Abstract*

We report the results of a double-slit-like experiment in the infrared range, which confirm those of a previous one by evidencing an anomalous behaviour of photon systems under particular (energy and space) constraints. These outcomes (independently confirmed by crossing photon beam experiments in both the optical and the microwave range) apparently rule out the Copenhagen interpretation of the quantum wave, i.e. the probability wave, by admitting an interpretation in terms of the Einstein-de Broglie-Bohm hollow wave for photons. Moreover, this second experiment further supports the interpretation of the hollow wave as a deformation of the Minkowski space-time geometry.

# 1. Introduction

## 1.1. The first experiment: Quantum wave and space-time deformation

In this paper, we report the results of an optical experiment (in the infrared range), which replicates a previous one [1], aimed at evidencing a possible anomalous behaviour of photon systems. Let us briefly report the main features and results of this first experiment, carried out in 2003 [1].

The employed apparatus (schematically depicted in Fig.1) consisted of a Plexiglas box with wooden base and lid. The box (thoroughly screened from those frequencies which might have affected the measurements) contained two identical infrared (IR) LEDs, as (incoherent) sources of light, and three identical photodiodes, as detectors. The two sources were placed in front of a screen with three circular apertures on it. Two of them were lined up with the two LEDs, so that each IR beam propagated perpendicularly through each of them. The geometry of this equipment was designed so that no photon could pass through the third aperture on the screen.

---

[*] Corresponding author.
 *E-mail address:* mignani@fis.uniroma3.it (R. Mignani).

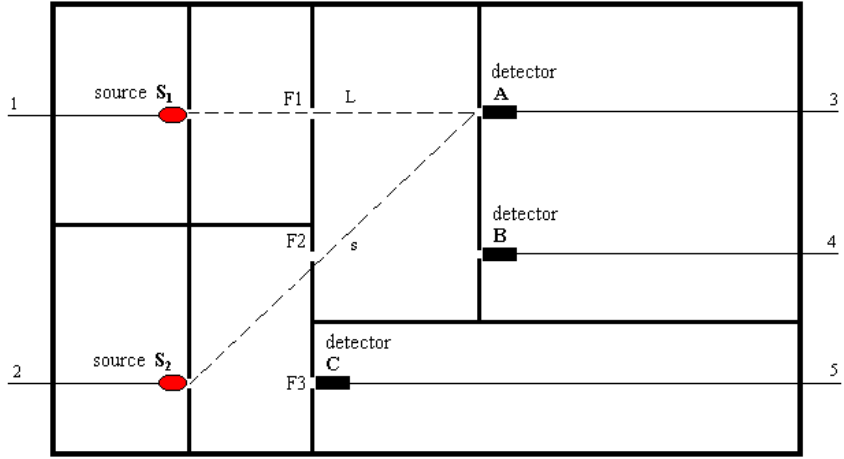

Fig. 1

Let us highlight the role played by the three detectors. Detector C destroyed the eigenstates of the photons emitted by $S_2$. Detector B ensured that no photon passed through the aperture $F_2$. Finally, detector A measured the photon signal from the source $S_1$.

In essence, the experiment just consisted in the measurement of the signal of detector A (aligned with the source $S_1$) in the two following conditions: (1) only the source $S_1$ switched on; (2) both sources on. Due to the geometry of the apparatus, *no difference in signal on A between these two conditions ought to be observed, according to either classical or quantum electrodynamics.* Therefore, a non-zero difference between these two values was considered evidence for the searched anomalous effect.

The outcome of this first experiment was positive. The envisaged effect was observed indeed [1].

As stressed in [1], such an anomalous behaviour for a photon system cannot be explained in the framework of the Copenhagen interpretation. On the contrary, it can be understood in terms of an interaction of photons with the Einstein-de Broglie-Bohm hollow waves belonging to those photons absorbed by detector C.

Moreover, the phenomenon exhibited a marked threshold behaviour. In fact, it was observed within a distance of at most 4 *cm* from the sources, and the measured signal difference on detector A ranged from $2.2 \pm 0.4$ $\mu V$ to $2.3 \pm 0.5$ $\mu V$ [1]. These values are consistent with the threshold behaviour for the electromagnetic breakdown of local Lorentz invariance (LLI), obtained by two of the present authors (F.C. and R.M) in the framework of the so-called Deformed Special Relativity (DSR) (i.e. a generalization of Special Relativity based on a "deformation" of the Minkowski space, with a metric whose coefficients depend on the energy of the investigated processes) [2][1].

Therefore, our first experiment allowed us to envisage a connection between the quantum wave (according to the Einstein-De Broglie-Bohm interpretation) and the breakdown of local Lorentz invariance (described by the DSR formalism). Namely, we hypothesized [1] that *a hollow wave is nothing but a deformation of space-time geometry*. By a metaphoric image we may picture the deformed space-time, which is intimately bound to each photon, as *the shadow of the photon*. It is immaterial, like a shadow (since it carries neither energy nor momentum) and it can fill space regions far from the photon, exactly as a shadow fills space regions far from the body that casts it. The space-time deformation spreads beyond the border of space and time sizes corresponding to the photon wavelength and period, respectively. This changes the photon-photon cross section and therefore similar effects might be observed in photon-photon interactions, for instance in crossing photon beams.

---

[1] More precisely, the analysis of the Cologne experiment [3] (superluminal sub-cutoff propagation in waveguide), and of the Florence one [4] (superluminal propagation in air), carried out by the DSR formalism, brought about upper threshold values both in energy and in space for the electromagnetic breakdown of LLI. These values are $E_0 = 4.5$ $\mu V$ and $l_0 = 9$ cm [2,5] respectively.

*1.2. Crossing photon beam experiments*

Further evidence for the anomalous photon system behaviour and for the anomalous photon-photon cross section was provided indeed by experiments with orthogonal crossing photon beams. These interference experiments have been carried out in the last year, one with microwaves emitted by horn antennas (see Fig.2), at IFAC – CNR (Ranfagni and coworkers) [6, 7], and the other with infrared $CO_2$ laser beams (Fig.3), at INOA – CNR (Meucci and coworkers) [7]. Let us summarize the results obtained.

The main result of the IFAC - CNR experiment consists in an unexpected transfer of modulation from one beam to the other, which cannot be accounted for by a simple interference effect. This confirms the presence of an anomalous behavior in photon systems, in the microwave range too.

Preliminary results of the optical experiment carried out at INOA-CNR have been reported in [1,7]. The wavelength of the used infrared laser beams was 10600 *nm*, namely one order of magnitude higher than the wavelength of the sources (LEDs) used in our experiments (850 *nm*).
The optimum alignment which can be achieved with lasers and the laser beam confinement make this optical set-up especially suitable for investigating the anomalous behaviour of the photon-photon cross section. The measurements were carried out for a relatively long lapse of time, that is, 12 minutes. This allowed one to perform a statistical test on the averaged results [7]. The signal statistics provided a significant variation in the mean values obtained with or without beam crossing. Hence the chance to have two identical statistics was rejected with a sufficient level of confidence. Our analysis of INOA experimental data is given in Fig. 4, where we highlighted the displacement of the crossed beam signal (in red) with respect to the single beam signal (in black). We concluded, after a deep statistical analysis of those two signals, that the actual shift is 2.08 ± 0.13 *µV*. This value agrees excellently with that obtained in our first experiment, which lies between 2.2 ± 0.4 *µV* and 2.3 ± 0.5 *µV*. Notice that *the laser experiment shows that the observed phenomenon does not depend either on the infrared wavelength, or on the coherence properties of the light.*

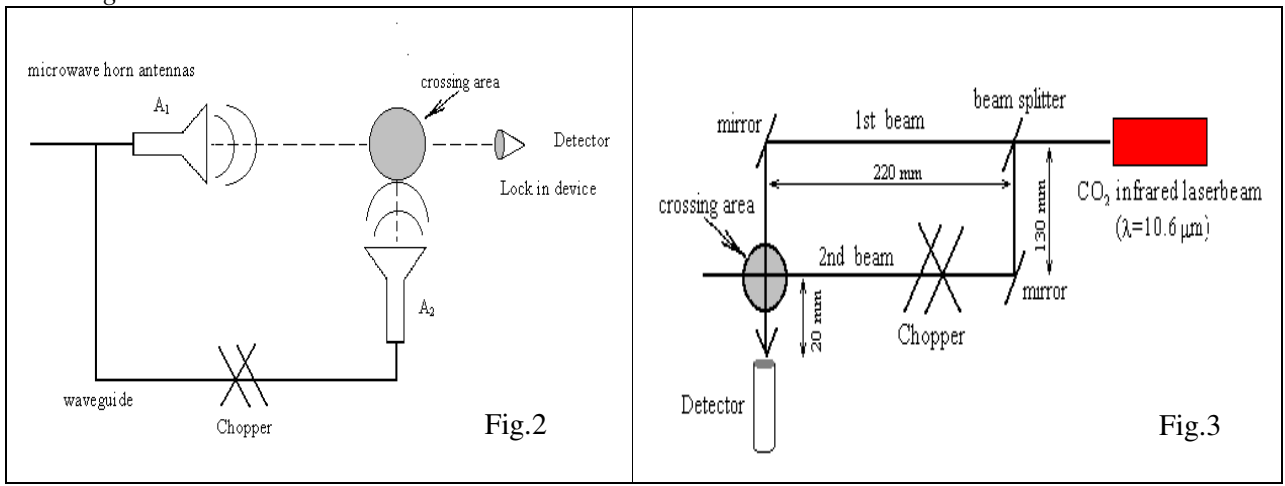

Fig.2    Fig.3

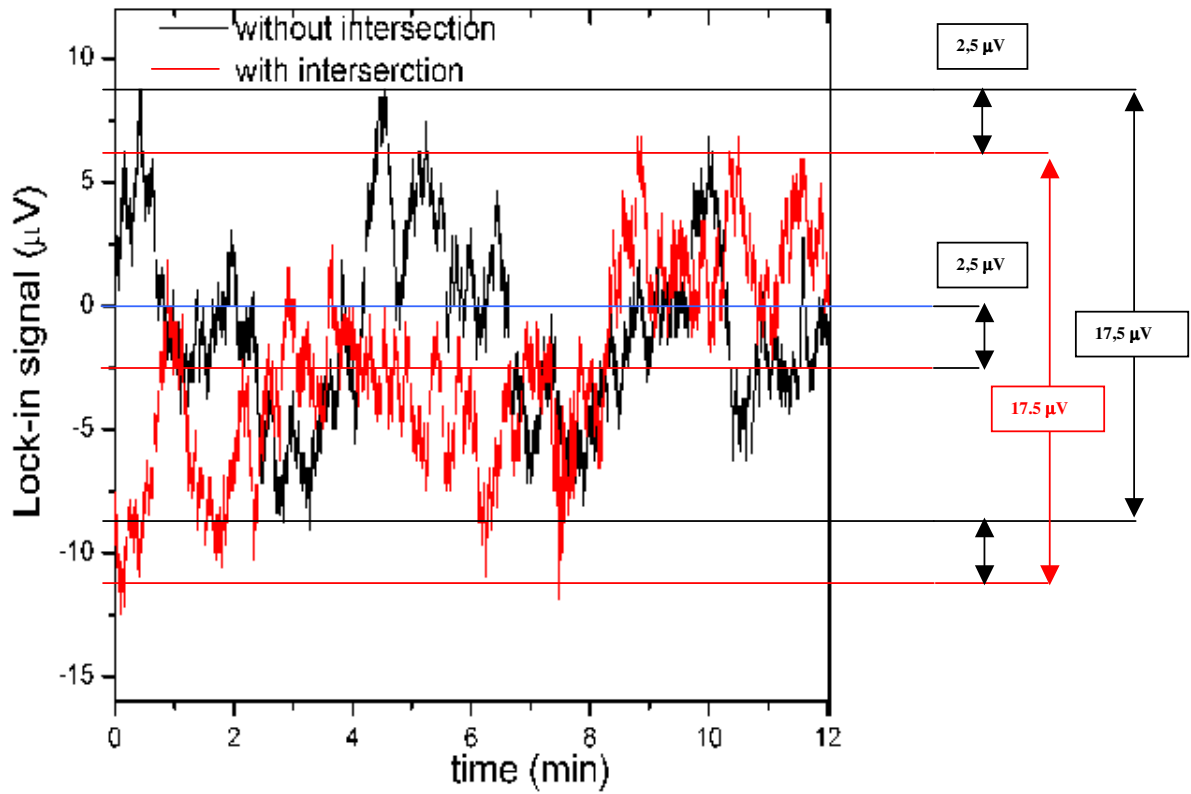

Fig. 4

## 2. The experiment

### 2.1. *Experimental set-up*

The purpose of the experiment reported in this paper was to corroborate the results of the previous one. The experimental set-up was essentially the same, but with a right-to-left inversion along the bigger side of the box (see Fig. 5) and with three detectors of a type different from that used in the first investigation. In this way, it was possible to study how the phenomenon changes under a spatial parity inversion[2] and for a different type of detectors.

The experimental equipment comprised a Plexiglas box with wooden base and lid, which contained two similar photon sources, three circular apertures and three similar phototransistors with convergent lenses. The layout of the experimental set-up, seen from above, is shown in Fig.5 and is, of course, the mirror image of Fig. 1.

The dimensions of the apparatus were identical to those of the first experiment. Let us recall that they were inferred from the geometrical size of the IFAC microwave experiment [4], namely the horizontal distance between the planes of the antennas.

The present experiment, which was performed at the University of L'Aquila during 2004, exploited two infrared radiation sources with a wavelength $\lambda = 8.5 \cdot 10^{-5}$ *cm*. The three circular apertures had a diameter of 0.5 *cm* and, being much greater than the wavelength $\lambda$, did not produce any diffraction phenomenon (except for the Fraunhofer diffraction, which was taken into account in the background measurements).

---

[2] Testing the phenomenon under parity was suggested to us by Tullio Regge (private communication).

The energy of the photons of the present and the previous experiment was $10^4$ times higher than that of the photons in the Cologne and Florence experiments [3,4,6], and 10 times higher than that of the INOA-CNR experiment [7]. Because of the unlike efficiency of the detectors employed in the two experiments, we expected to measure a signal difference on detector A, due to the breakdown of LLI, unlike the signal difference measured in the first experiment. The three identical detectors were phototransistors of the type with a convergent lens. Both their bias signal and their photo-currents were conveyed to them and from them by electric wires which passed through three separate holes on the box side, as shown in Fig. 5. The wires were completely screened from those electrical frequencies involved in the experiment and so were the holes, with respect to optical frequencies of course. In this way we could detach the experimental equipment, i.e. the box and its content, both from the voltage generator and the measuring tools.

The Plexiglas walls of the box were 0.4 *cm* thick, with their outside surfaces covered by a black film. When the lid was put on the box, no external disturbances, due to radiation with the frequencies involved in the experiment, could filter through, as it was checked out explicitly by measuring the dark voltage stability.

Detector C was fixed in front of the source $S_2$; detectors A and B were fixed on a common vertical panel (see Fig. 5).

By assuming the existence of the de Broglie-Bohm wave connected to a photon, the photons from the source $S_1$ interact with the shadows of photons from the source $S_2$, which have gone through the aperture $F_2$. Consequently, detector A measures a signal unlike the one detected when $S_1$ is turned on and $S_2$ is turned off.

In this new version of the experiment, the measurements were carried out only for the distance *L* (between the plane of the sources and the plane of the detectors A and B, see Fig. 5) equal to 1*cm*, unlike the former investigation. The reason for this choice will be provided later on.

The role played by the aperture $F_2$ is fundamental, since, if we assume the wave to be a space-time deformation, the encountered mass density affects its propagation. Hence, it can pass only through space regions with a low mass density.

Due to the role played by the three detectors (in analogy with the first experiment: see Subsect.1.1), if detectors B and C do not manifest any change in their response signal, any variation in the signal on detector A can be attributed *only* to the interaction of the photons emitted by the source $S_1$ with the hollow waves passed through the aperture $F_2$.

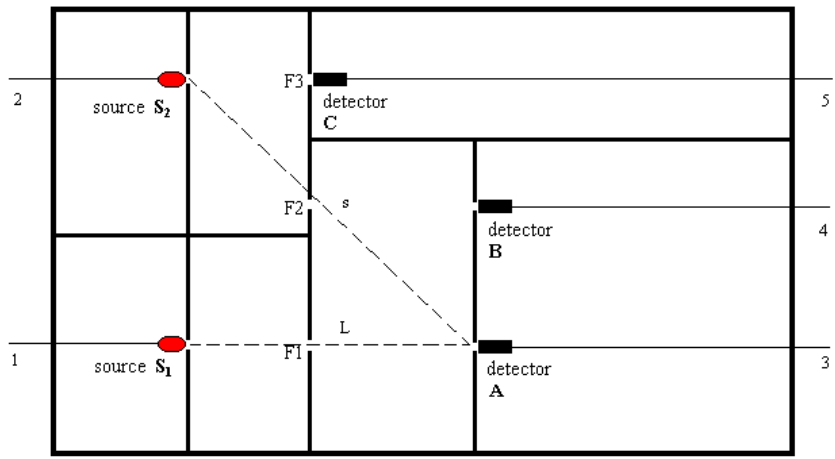

Fig. 5

*2.2. Measurement conditions*

The two employed sources were two identical infrared LEDs of the kind High Speed Infrared Emitter AlGaAs (HIRL 5010, Hero Electronics Ltd.), with an emission peak at 850 *nm* and angular aperture of 20°. About the 65% of the radiated intensity ranged from 840 to 870 *nm*.
The three detectors were planar epitaxial silicon NPN phototransistor in a case 18 A 3 DIN 41876 (TO 18) (BpY 62/III, Siemens). They were employed with an open base configuration, and had a maximum sensitivity wavelength (photo-current /unit surface) equal to 800 *nm*. Their response, as a function of the angle between the incidence direction of the radiation and the normal to the sensitive surface, was zero for angles wider than about 10°. Within the range 600 – 900 *nm*, the sensitivity was greater than 60% of the maximum, and the emission peak of the LEDs was 90% of the maximum. This ensured a good alignment of the detectors with the axis of the radiation lobe of the sources and maximum efficiency tuning of the detectors on the received wavelengths. The phototransistor sensitive surface had an area of 0.14 $mm^2$.
Since it was crucial to have the same working conditions for both sources and detectors, two separate power supplies were used, one for the former and one for the latter. The LEDs and the phototransistors were fed in parallel by two identical constant voltage generators (mod. Labornetzgeraet/regulated power supply LAB 510).
The measurements were taken by means of a digital multimeter (mod. Agilent 34401 A), with 6 ½ digit resolution (overrange) [8].

We initially checked that, when the supply voltage of the sources was held constant, the response voltage of the detectors was stable and linear with the drain-source voltage $V_{ds}$, when this was kept in a range from 1.5 to 5 *V*. Hence, the drain-source voltage $V_{ds}$ was fixed at 3.00 *V*. Analogously to the previous experiment, we operated as closer as possible to the few-photon condition[3], namely, with the lowest radiated intensity from the LEDs which would induce a detector response distinguishable from the dark signal. This was accomplished by regularly increasing the supply voltage of the source $S_1$. We started from zero voltage until a response voltage of 7$\mu V$ was measured on detector A. The corresponding supply voltage (the same for the two sources), which yielded 7$\mu V$ on A, surely generated a greater signal on detector C (since C was nearer to the corresponding source $S_2$). This signal was easier to distinguish from the dark, as it was experimentally ascertained.

We expected to find a lower signal difference on A, connected to the breakdown of LLI, compared with that measured in the first experiment. This is due to the fact that the detectors in this experiment (phototransistors) had a lower "relative efficiency" compared with the detectors (photodiodes) used in the previous investigation. The detection sensitive area and the slew rate (detection time from 10% to 90% of the final response signal) were different in the two types of detectors (see Table 1).

**Table 1**

**Parameters of the detector response (from the Data Sheet)**

| Experiment | Detector | Sensitive Area ($mm^2$) | Detection time ($\mu s$) |
|---|---|---|---|
| 1° | Photodiode | 5.244 | 90 |
| 2° | Phototransistor | 0.140 | 5 |

Then, it is worth defining the relative geometrical efficiency ($\eta_g$) of the phototransistor, with respect to the photodiode, as the ratio of their respective sensitive areas, and their relative time efficiency

---

[3] At least with respect to the optimum conditions of maximum luminosity suggested by the manifacturer.

($\eta_t$) as the ratio of their respective detection times. Likewise, one can define the relative total efficiency ($\eta_T$) of the phototransistor with respect to the photodiode as the product of the two efficiencies defined above.
The values of these three efficiency parameters are given in Table 2.

**Table 2**

**Relative efficiencies of the detector employed in the present experiment**

| $\eta_g$ | $\eta_t$ | $\eta_T$ |
|---|---|---|
| 0.0267 | 0.0556 | 0.0015 |

Therefore, it was reasonable to presume that the value of the expected phenomenon to be given by the product of the total relative efficiency times the value measured in the first experiment, i.e. $\eta_T \cdot (2.3 \pm 0.5 \mu V) = 0.004 \pm 0.001~\mu V$. This was the difference in signal expected to be measured by detector A between the two source states, (1) $S_1$ on, $S_2$ off; (2) both sources on. This predicted value is two orders of magnitude below that measured in the first experiment. Therefore, we foresaw that the phenomenon would have been observable only at distance of 1 *cm* (which corresponds to the position number 1 in the previous experiment, when we investigated also the spatial extension of the effect: see ref. [1]).

### 2.3. *Measurement procedure*

The measurement procedure was designed in order to avoid unclear situations that arise when the real phenomenon is simulated by variations in the supply voltage during the measurement. This is why we initially tested – and we did it again during every single measurement – the stability of the supply voltage both for the sources and for the detectors. It was found to be stable in an interval less than 1 *mV*. We checked, for each detector, that this voltage variation did not induce any change in the multimeter reading within its corresponding error.
Each group of measurements was a collection of single measurements carried out according to the following procedure:

**Step 1:** Measurement of the signal from detector A with source $S_1$ turned on and source $S_2$ turned off;
**Step 2:** Measurement of the signal from detectors A, B and C with $S_1$ off and $S_2$ on;
**Step 3:** Measurement of the signal from A, B and C with both sources $S_1$ and $S_2$ on.

Since the phototransistor detection time is 5 $\mu s$, we chose 5 *s* as time interval for data sampling. We checked that over a double time interval, i.e. 10 *s*, the maximum measured value of the dark voltage for the three phototransistors was 0.7 $\mu V$. This value was interpreted as the maximum pessimistic error. We adjusted the multimeter in order to sample the voltage 6 times per second. During every single sampling interval (5 *s*), the maximum and the minimum of the $6 \cdot 5 = 30$ samples (gaussian samples) were recorded in the multimeter memory.
Hence, every measurement step yielded 2 values for each detector, the minimum and the maximum one. Thus an entire cycle of three steps yielded $2 \cdot 7 = 14$ samples, 7 minimum values and 7 maximum values. We tested the performances of the multimeter automatic sampling, reported in the handbook, by carrying out our own sampling test and verifying the reported error. Every single minimum and maximum value was affected by a 0.003 $\mu V$ error (thanks to the chosen sampling time distribution), which agrees with the precision performances of the multimeter within the measurement adopted conditions.

## 2.4. Results.

Thirty groups of measurements were gathered (gaussian samples), and, by applying an inferential statistical test, we obtained the gaussian distribution of the minimum and maximum values for every measurement and each detector. The expectations of these two distributions (minimum values, maximum values) were computed for each detector. Then, we calculated the mean of these two expectation values (total mean $\overline{V}$). Let us denote by $\Delta \overline{V}$ the difference of the total averages, corresponding to the measurements on each detector for the two states of the sources. The experimental results are reported in Table 3.

**Table 3**

**Experimental results**

| Sources | TOTAL MEAN $\overline{V}$ ($\mu V$) | | | | | | DIFFERENCES BETWEEN TOTAL MEANS | |
|---|---|---|---|---|---|---|---|---|
| | $S_1$ | $S_2$ | $S_1$ | $S_2$ | $S_1$ | $S_2$ | SIGN | ABSOLUTE VALUE $\Delta \overline{V}$ ($\mu V$) |
| State | ON | ON | ON | OFF | OFF | ON | | |
| Detector A | 8.634±0.003 | | 8.626±0.003 | | | | (ON ON) > (ON OFF) | 0.008±0.003 |
| Detector B | 0.020±0.003 | | | | | 0.275±0.003 | (ON ON) < (OFF ON) | 0.255±0.003 |
| Detector C | 16.226±0.003 | | | | | 16.476±0.003 | (ON ON) < (OFF ON) | 0.250±0.003 |

The differences between the total means of the measurement on each detector for the two possible states of the sources allow us to draw the following conclusions.
The detector B was always underneath the maximum dark threshold that corresponds to 0.7 $\mu V$. The value of the difference for B had the same sign and the same order of magnitude of that of the detector C, which, conversely, was always exposed to radiation. Hence, we could speak of a common difference for the detectors B and C, which can be regarded as a *device* signal effect. The difference for detector A had an opposite sign with respect to that of detectors B and C and was lower by two orders of magnitude.
Since the detector B was always below the maximum dark threshold, it can be inferred that no photons from $S_2$ went through the aperture $F_2$[4]. Therefore, the disparity between the difference on detector A and those on detectors B and C *cannot be attributed to photons that passed through the aperture $F_2$*. As a consequence, we regard the difference on A *as a true signal difference* and not as a device effect.

---
[4] In order to support this statement, we point out that the total mean of the signal on detector B when both sources are on is lower of one order of magnitude than that when only S2 is on. Since, between the two sources, S1 can affect more the response of B and switching it on makes decrease the total mean of B, we can infer that S2 affects it much less and hence no photons from it can go through F2.

Moreover, the value of the difference measured on detector A (0.008 ± 0.003 $\mu V$), is consistent, within the error, with the difference 2.3 $\mu V$ measured in the first experiment, *provided that the unlike efficiency between photodiodes and phototransistors is taken into account* (see Subsect.2.2 and Tables 1, 2). Let us in fact recall that the expected value, on account of the total relative efficiency between phototransistors and photodiodes reported in Table 2, was just estimated to be $\eta_T \cdot (2.3 \pm 0.5 \ \mu V) = 0.004 \pm 0.001 \ \mu V$.

## 3. Conclusions and remarks

We believe that the outcomes of this second experiment confirm the existence of the anomalous behaviour in photon interference for the three following reasons:

i) Consistency of the measured value with that obtained in the first experiment, despite the new type of detectors employed;
ii) The sign of the signal variation on the detector A (that measures the phenomenon), which is always opposite to the signal variations on the two controlling detectors B, C;
iii) The difference of about two orders of magnitude between the variation of the signal on A and the variation common to B and C.

Moreover, the observed effect is apparently not affected by the parity of the equipment.

Let us stress again that our experimental results do seemingly favour the Einstein-de Broglie-Bohm interpretation of quantum wave with respect to the Copenhagen one[5] and, moreover, they seem to be consistent with our picture of the hollow wave as a space-time deformation.

Actually, by recalling the role played by the three detectors in both experiments (see Subsects. 1.1, 2.2), since C measures – and hence destroys – the superposition of states belonging to the photons emitted by $S_2$ (thus manifesting their corpuscle nature) and B ensures no transit of these photons through aperture $F_2$, there is no way the Copenhagen interpretation can explain the signal difference (0.008 $\mu V$) measured by detector A.

Conversely, our hypothesis about the hollow wave – regarded as a space-time geometry deformation intimately bound to the quantum entity – succeeds in accounting for such a difference. This hypothesis can be briefly depicted as follows. Most of the energy of the photon is concentrated in a tiny extent; the remaining part is employed to deform the space-time surrounding it and, hence, it is stored in this deformation. It is just the deformations (the "shadows") of the photons from $S_2$ that expand, go through $F_2$ and interact with the shadows of the photons from $S_1$.

Therefore, in our opinion, the difference of signal measured by the detector A in both our experiments can be interpreted *as the energy absorbed by the space-time deformation itself*, which cannot be detected by the central detector B[6]. In other words, our experimental device, used in both experiments, "weighed" the energy corresponding to the space-time deformation by the measured difference on the first detector.

In this connection, let us notice that the hollow wave seen as "shadow of light" (which affects quantum objects in seemingly inaccessible and far regions) apparently represents an action-at-a-distance without any transport of energy (which Einstein, within the domain of Quantum Mechanics, called "spooky action at a distance"). However, just our interpretation of the hollow wave as a space-time deformation which moves together with the quantum object – the photon in this case – is actually an indication of the opposite view. As a matter of fact – as discussed above –, part of the photon energy is detected by a direct measurement of photons by the third detector C; the remaining part is used to deform the space-time of every photon and it is evidenced by the difference measured by the first detector. Hence, it is no longer correct to say that there exists (at least in this framework) an action-at-a-distance without any energy transport.

---

[5] Actually, the probabilistic interpretation of the wave-function was due to M. Born of the Göttingen school
[6] One might think to detect such an "energy of deformation field" (corresponding to the hollow waves of photons) by a detector operating by the gravitational interaction, rather than the electromagnetic one. However, this would still be impossible, because the deformation value lies within the energy interval for a flat (Minkowski) gravitational space-time, according to DSR [2]. We are deeply indebted to G. Caricato for this and other precious remarks on the topics of action-at-a-distance in our experiments (correspondence between Caricato and F. Pistella).

If the interpretation we have given here is correct, *our experiments, among the others, do provide for the first time direct evidence for the Einstein-de Broglie-Bohm waves and yield a measurement of the energy associated to them.*

*Acknowledgements*

It is both a pleasure and a duty to warmly thank Gaetano Caricato and Tullio Regge, to whom we are strongly indebted for their deep review and highly useful remarks. We are also grateful to Riccardo Meucci for communicating to us the preliminary results of his experiment. Stimulating discussions with (and useful criticism by) Basil Hiley, Eliano Pessa and Anedio Ranfagni are gratefully acknowledged. Last but not least, a special thank is due to the President of CNR, Fabio Pistella, for his kind interest in our work and continuous encouragement.

*References*

[1]  F. Cardone, R. Mignani, W. Perconti, R. Scrimaglio, *Atti della Fondazione Giorgio Ronchi* **LVIII** (2003) 869; *Phys. Lett. A* **326** (2004) 1.

[2]  F. Cardone, R. Mignani, *Energy and Geometry – An Introduction to Deformed Special Relativity* (World Scientific, Singapore, 2004).

[3]  G. Nimtz, A. Enders, H. Spieker, *J. Phys. I (Paris)* **4** (1994) 1; W. Heitmann, G. Nimtz, *Phys. Lett. A* **196** (1994) 154.

[4]  A. Ranfagni, P. Fabeni, G. P. Pazzi, D. Mugnai, *Phys. Rev. E* **48** (1993) 1453.

[5]  F. Cardone, R. Mignani, *Phys. Lett. A* **306** (2003) 265.

[6]  A. Ranfagni, D. Mugnai, R. Ruggeri, *Phys. Rev. E* **69** (2004) 027601; A. Ranfagni, D. Mugnai, *Phys. Lett. A* **322** (2004) 146.

[7] D. Mugnai, A. Ranfagni, E. Allaria, R. Meucci, C. Ranfagni, "The question of the superluminal speed of information: Unexpected behavior in the crossing of microwave and optical beams", submitted for publication (2004).

[8]  Agilent Technologies, Multimeter 34401A, Work Handbook, p. 21.

**FIGURE CAPTIONS**

**Fig. 1** -  Above view of the experimental apparatus used in the first experiment.

**Fig. 2** – Schematic view of the crossed-beam experiment in the microwave range, exploiting two horn antennas.

**Fig. 3 –** Schematic view of the crossed-beam experiment in the infrared range, exploiting a $CO_2$ laser emitting at 10.6 $\mu m$ on the fundamental $TEM_{00}$ Gaussian mode. The laser beam is split in two orthogonal beams (beam 1 and beam 2) by means of a beam splitter. By using two flat mirrors the two beams are directed to the crossing area within the near field of the Gaussian mode estimated at 1.5 *m* from the out-coupler mirror of the laser cavity. Beam 2 is periodically interrupted by means of a chopper whose frequency is the reference frequency in a lock-in amplifier connected to the detector.

**Fig. 4** - Output signal of the lock-in amplifier in the presence of the chopped beam 2 (red line) and in its absence (black line).

**Fig. 5** -  Above view of the experimental apparatus used in the present experiment; note that it is the mirrored image of Fig. 1.